\documentclass[12pt,letterpaper]{article}
\pdfoutput=1

\usepackage[dvips]{epsfig}
\usepackage{color}
\definecolor{refkey}{rgb}{1,0,0}
\definecolor{labelkey}{rgb}{0,0,1}
\usepackage{amsfonts,amssymb,amsmath}
\usepackage{float}
\usepackage{caption}
\usepackage{subcaption}
\usepackage{rotating}
\usepackage{appendix}

\newcommand{\be}{\begin{equation}}
\newcommand{\ee}{\end{equation}}
\newcommand{\ben}{\begin{displaymath}}
\newcommand{\een}{\end{displaymath}}
\newcommand{\bean}{\begin{eqnarray*}}
\newcommand{\eean}{\end{eqnarray*}}

\newcommand{\ads}[1]{\mbox{${AdS}_{#1}$}}

\newcommand{\eg}{{\it e.g.}}

\newcommand{\commentout}[1]{}

\renewcommand{\theequation}{\arabic{section}.\arabic{equation}}

\newcommand{\beq}{\begin{equation}}
\newcommand{\eeq}{\end{equation}}
\newcommand{\beqr}{\begin{displaymath}}
\newcommand{\eeqr}{\end{displaymath}}
\newcommand{\beqa}{\begin{eqnarray}}
\newcommand{\eeqa}{\end{eqnarray}}
\newcommand{\beqar}{\begin{eqnarray*}}
\newcommand{\eeqar}{\end{eqnarray*}}
\newcommand{\cN}{{\cal N}}

\newcommand{\cA}{{\cal A}}

\newcommand{\N}[1]{\ensuremath{\cN=#1}}

\begin{document}

\title{\LARGE \bf Minimal area surfaces in \ads{3} through integrability}

\author{
	 Yifei He\thanks{E-mail: \texttt{he163@purdue.edu}} ,
	Martin Kruczenski\thanks{E-mail: \texttt{markru@purdue.edu}} \\
	Department of Physics and Astronomy, Purdue University,  \\
	525 Northwestern Avenue, W. Lafayette, IN 47907-2036.}

\maketitle

\begin{abstract}
 Minimal area surfaces in \ads{3} ending on a given curve at the boundary are dual to planar Wilson loops in \N{4} SYM. In previous work 
 it was shown that the problem of finding such surfaces can be recast as the one of finding an appropriate parameterization of the boundary contour
 that corresponds to conformal gauge. A. Dekel was able to find such reparameterization in a perturbative expansion around a circular contour.
 In this work we show that for more general contours such reparameterization can be found using a numerical procedure that does not rely on a perturbative expansion. 
 This provides further checks and applications of the integrability method. An interesting property of the method is that it uses as data the Schwarzian derivative of the contour
 and therefore it has manifest global conformal invariance. Finally, we apply Shanks transformation to extend the near circular expansion 
 to larger deformations. The results are in agreement with the new method. 
 
\end{abstract}

\clearpage
\newpage

\section{Introduction}
According to the AdS/CFT correspondence \cite{Maldacena:1997re,Gubser:1998bc,Witten:1998qj}, the expectation value of the Wilson loop in $SU(N )$ \N{4} SYM theory, for large $N$ and at large 't Hooft coupling can be computed by finding a minimal surface in AdS space \cite{Maldacena:1998im,Rey:1998ik} ending at the boundary on the Wilson loop. In the case of \ads{3}, a standard method to find such surfaces is through Pohlmeyer reduction \cite{Pohlmeyer}. The equation of motion is simplified to a linear problem accompanied by a generalized cosh-Gordon/sinh-Gordon equation. Once the equation is solved, one can construct the surface and calculate the area. Over the years, much work have been done on the computation of Wilson loops of various shapes. In Minkowski signature, the most interesting cases are Wilson loops with light-like cusps \cite{Kruczenski:2002fb} due to their relation with scattering amplitudes \cite{Alday:2007hr,Alday:2009yn}. In Euclidean signature, the well-studied cases include circular Wilson loops \cite{Berenstein:1998ij}, the wavy Wilson loops \cite{Semenoff:2004qr}, the cusp \cite{DGO} and more generally solutions in terms of Riemann theta functions \cite{Ishizeki:2011bf,Kruczenski:2013bsa}. Although the Pohlmeyer reduction allows to find solutions, in general, given an arbitrary smooth contour, it is not known how to find the minimal surface ending on it and compute the area. Essentially, the complication is that we have to solve an elliptic problem for an integrable system instead of a time evolution problem as is more common. In this paper, we focus on this problem. We consider the Euclidean case, namely, an Euclidean Wilson loop confined on a plane such that the dual surface is contained on a $\mathbb{H}_{3}$ subspace of $AdS_5$. 

A formalism for approaching this problem in the Euclidean case was recently introduced in \cite{Kruczenski:2014bla} where the calculation of the area of the minimal surface ending on a given boundary contour was reduced to finding a parameterization of the contour in terms of the conformal angle $\theta$ on the corresponding worldsheet. Once the conformal angle is found, one can express the area in terms of the Schwarzian derivative of the boundary contour with respect to the conformal angle. This formalism was used to study contours perturbatively around circular contours by A. Dekel in \cite{Dekel:2015bla}, where the area was given as a series expansion in the perturbative parameter to high order. In \cite{Irrgang:2015txa}, the method was generalized to Minkowski case and in \cite{Huang:2016atz}, solutions given by Mathieu functions were found. However, a general analytical or numerical solution to the problem of finding the conformal parameter for a given contour is not known. In this paper we use the formalism given in \cite{Kruczenski:2014bla,Dekel:2015bla} and provide a numerical solution to the problem. Our main objective is to provide a check and an application of the integrability ideas that were used to develop the method. In particular, an important aspect of integrability that is manifest in this method is the existence of a one parameter family of curves with the same area related by a symmetry that changes the spectral parameter \cite{Ishizeki:2011bf,Kruczenski:2014bla} known as $\lambda$-deformations \cite{Dekel:2015bla} or "master" symmetry \cite{Klose:2016uur,Klose:2016qfv}. In fact, in this last work it was shown that such symmetry can be used to construct the non-local Yangian charges from the global symmetries. 
 
The paper is organized as follows. In section \ref{setup}, we give a brief review of the general setup and previous results for studying minimal surfaces in $\mathbb{H}_{3}$. Following the formalism in \cite{Kruczenski:2014bla}, we describe how the boundary condition of the cosh-Gordon equation is encoded in the boundary contour and how the problem is reduced to finding the correct parameterization of the boundary contour. In the following section, we describe the numerical method used to find such parameterization and give examples for
various contours. In section \ref{shanks}, we extend the perturbative results given in \cite{Dekel:2015bla} to regions where the original expansion diverges and reproduce the results of the new method as a check. In section \ref{areaforzeros}, we provide an area formula for contours where the Pohlmeyer holomorphic function $f(z)$ has zeros which the area formula given in \cite{Kruczenski:2014bla} fails to apply to. The last section gives our conclusions. It should be noted that it is also possible to attempt to solve the minimal surface directly, see \eg\ \cite{FT} and more recently \cite{Klose:2016uur,Klose:2016qfv} where the $\lambda$-deformations were also constructed numerically. Here we concentrate in understanding the integrability properties of the system and use the numerical solutions as a check of the integrability ideas. Also it should be noted that, although our numerical method is general, in practice it converges slowly if the contour is irregular and cannot be described accurately by interpolating through a relatively small set of points. 

\section{General setup}\label{setup}
In this section, we review the general setup for studying minimal surfaces in $\mathbb{H}_{3}$. We briefly describe the method given in \cite{Kruczenski:2014bla} which reduces the problem of calculating the area to finding the conformal parametrization and explain how it was used in \cite{Dekel:2015bla} to find solutions perturbatively around the circular contour. 
\subsection{Minimal surfaces in $\mathbb{H}_{3}$}
In $\mathbb{R}^{1,3}$, $\mathbb{H}_{3}$ is embedded as a hyperboloid $X\cdot X=-1$, where $X=(X_{0},X_{1},X_{2},X_{3})$. The metric in $\mathbb{R}^{1,3}$ is
\begin{equation}
ds^2=-dX_0^2+dX_1^2+dX_2^2+dX_3^2.
\end{equation}
The Poincar\'{e} coordinates are given by
\begin{equation}
Z=\frac{1}{X_{0}-X_{3}}, \quad X=\frac{X_{1}+iX_{2}}{X_{0}-X_{3}}, \quad \bar{X}=\frac{X_{1}-iX_{2}}{X_{0}-X_{3}}.
\end{equation}
and the metric is
\begin{equation}
ds^2=\frac{dZ^2+dXd\bar{X}}{Z^2}.
\end{equation}

An Euclidean surface in $\mathbb{H}_{3}$ can be described as a map $X(r,\theta)$, $Z(r,\theta)$ from the unit disk on the complex plane parameterized as $z=re^{i \theta}$ ($r\le1$) and we assume
a conformal parameterization, namely the induced metric is
\beq
 ds^2 = 4 e^{2\alpha}\, dz\,d\bar{z}
 \label{indmetric}
\eeq 
for some real function $\alpha(z,\bar{z})$. As $r\to 1$, one approaches the boundary of the surface, where
\begin{equation}
Z(r=1,\theta)=0, \quad X(r=1,\theta)=X(s(\theta)).
\end{equation}
$X(s)$ is a given closed curve defined on the boundary of $\mathbb{H}_{3}$ with an arbitrary parameter $s$, which is related to the conformal angle $\theta$ by an unknown reparametrization $s(\theta)$.

The string action (area) is given by 
\begin{equation}
S=\frac{1}{2}\int{d\sigma d\tau (\partial X\cdot \bar{\partial} X+\Lambda(X\cdot X+1))},
\end{equation}
where $\Lambda$ is a Lagrange multiplier and we also need to impose the Virasoro constraints 
\begin{equation}
\bar{\partial} X\cdot\bar{\partial} X=\partial X\cdot\partial X=0.
\end{equation}
The equation of motion is
\begin{equation}
\partial\bar{\partial}X-\Lambda X=0,\ \ \  \ \Lambda=\bar{\partial} X\cdot\partial X.
\end{equation}

Using the equivalence $SO(1,3)\simeq SL(2,\mathbb{C})$, one can write
$\mathbb{X}= X_0+X_i\sigma^i$ where $\sigma^i$ are the Pauli matrices. The equation of motion and the Virasoro constraints become
\begin{equation}
\det\mathbb{X}=1, \quad \partial\bar{\partial}\mathbb{X}=\Lambda \mathbb{X}, \quad \det(\partial \mathbb{X})=\det(\bar{\partial}\mathbb{X})=0.
\end{equation}
 The matrix $\mathbb{X}$ satisfies the reality condition $\mathbb{X}^\dagger=\mathbb{X}$ that can be solved by writing
\begin{equation}
\mathbb{X}=\mathbb{AA}^{\dagger},
\end{equation}
with
\begin{equation}
\det\mathbb{A}=1, \quad \mathbb{A}\in SL(2,\mathbb{C}).
\end{equation}
 The matrix $\mathbb{A}$ satisfies the linear problem
\begin{equation}\label{linearequation}
\partial \mathbb{A}=\mathbb{A} J,\quad \bar{\partial} \mathbb{A}=\mathbb{A} \bar{J}, 
\end{equation}
where
\begin{equation}
J=\begin{pmatrix}
-\frac{1}{2}\partial\alpha&fe^{-\alpha}\\[6pt]
\lambda e^{\alpha}&\frac{1}{2}\partial\alpha
\end{pmatrix}, \quad 
\bar{J}=\begin{pmatrix}
\frac{1}{2}\bar{\partial}\alpha&\frac{1}{\lambda} e^{\alpha}\\[6pt]
-\bar{f}e^{-\alpha}&-\frac{1}{2}\bar{\partial}\alpha
\end{pmatrix}.
\end{equation}
Here $J,\bar{J}$ are the components of the current
\begin{equation}
j=\mathbb{A}^{-1}d\mathbb{A}=Jdz+\bar{J}d\bar{z}.
\end{equation}
The consistency condition requires $f$, $\bar{f}$ to be holomorphic and anti-holomorphic, and $\alpha(z,\bar{z})$ to satisfy the generalized cosh-Gordon equation:
\begin{equation}\label{coshGordon}
\partial\bar{\partial}\alpha=e^{2\alpha}+f\bar{f}e^{-2\alpha}.
\end{equation}
The expressions for $J$ and $\bar{J}$ include a spectral parameter $\lambda$. When $|\lambda|=1$ we obtain a one parameter family of minimal surfaces satisfying the equation of motion with different boundary contours but the same area. The $\lambda$-deformation of the original contour plays an important role in understanding the integrability of the problem and has been studied recently in \cite{Klose:2016uur,Klose:2016qfv}. In this paper we take $\lambda=1$, after obtaining such solution it is possible to change $\lambda$ and study the full $\lambda$-deformed family of contours but we leave that for future work.

 To use the above formalism, we have to find the function $f(z)$ associated with the particular contour $X(s)$ we are interested in, then solve the cosh-Gordon equation, write down the current $J,\bar{J}$ and solve for $\mathbb{A}$ from which we can reconstruct the minimal surface. To calculate the area, consider the induced metric \eqref{indmetric}, then the area is given by the integral
\begin{equation}
\cA=4\int_{D}e^{2\alpha}d\sigma d\tau.
\end{equation}
After regularization, the finite part of the area is (see \eg \cite{Kruczenski:2014bla})
\begin{equation}\label{regarea}
\cA_f=-2\pi-4\int_{D} f\bar{f}e^{-2\alpha}d\sigma d\tau,
\end{equation}
where $D$ is the unit disk on the complex plane.

\subsection{Boundary data}
Near the boundary, $r\to 1$ and it is convenient to define a world-sheet coordinate
\begin{equation}
\xi=1-r^2,
\end{equation}
Then, $\alpha(z,\bar{z})$ has the expansion
\begin{equation}\label{alphaexpand}
\alpha(\xi,\theta)\simeq-\ln\xi+\beta_2(\theta)(1+\xi)\xi^2+O(\xi^4),
\end{equation}
where $\beta_2(\theta)$ can be defined as
\begin{equation}
\beta_2(\theta)=\frac{1}{6}e^{2i\theta}(\partial^2\alpha-(\partial\alpha)^2)\big|_{r\to1}.
\end{equation}
All the higher order coefficients in \eqref{alphaexpand} are fixed by $\beta_2(\theta)$ and $f(\theta)=f(e^{i\theta})$.

Going back to the linear problem \eqref{linearequation} and writing $\mathbb{A}$ in terms of two linear independent vectors $\psi=(\psi_1,\psi_2)$ and $\tilde{\psi}=(\tilde{\psi}_1,\tilde{\psi}_2)$ as
\begin{equation}
\mathbb{A}=\left(\begin{array}{cc}
\psi_1&\psi_2\\
\tilde{\psi}_1&\tilde{\psi}_2
\end{array}\right),
\end{equation}
the linear equations \eqref{linearequation} are reduced to
\begin{equation}
\partial\psi=\psi J, \quad \bar{\partial}\psi=\psi\bar{J},
\end{equation}
and the same equations for $\tilde{\psi}$. Taking this linear problem to the boundary, it follows that \cite{Kruczenski:2014bla}
\begin{equation}\label{schwarzian}
\{X_\lambda(\theta),\theta\}=\frac{1}{2}-12\beta_{2}(\theta)-2\lambda f(\theta) e^{2i\theta}+\frac{2}{\lambda}\bar{f}(\theta)e^{-2i\theta},
\end{equation}
where $\{X_\lambda(\theta),\theta\}$ is the Schwarzian derivative of the boundary contour $X_\lambda(\theta)$ associated to a given value $\lambda$ of the spectral parameter. The original contour corresponds to $\lambda=1$, and one has 
\begin{equation}\label{schwarzianRI}
\begin{aligned}
&\text{Re}\{X(\theta), \theta\}=\frac{1}{2}-12\beta_2(\theta),\\
&\text{Im}\{X(\theta), \theta\}=-4\text{Im}(e^{2i\theta}f(\theta)).
\end{aligned}
\end{equation}
Hence, if we know the contour in terms of the conformal angle $\theta$, we can calculate the Schwarzian derivative with respect to $\theta$ and then obtain $\beta_{2}(\theta)$ and $f(\theta)$ from its real and imaginary parts. With such information, we can find out $f(z)$ by analytic continuation and plug it into the cosh-Gordon equation \eqref{coshGordon} to solve for $\alpha(z,\bar{z})$. Finally we can calculate the regularized area using \eqref{regarea}.

For a Wilson loop, the contour $X(s)$ is given in terms of an arbitrary parameter $s$ (instead of the conformal angle $\theta$). Using the property of the Schwarzian derivative
\begin{equation}\label{schwarzianproperty}
\{F,\theta\}=\{s,\theta\}+(\partial_{\theta}s)^2\{F,s\},
\end{equation}
eq.\eqref{schwarzianRI} gives
\begin{equation}
\begin{aligned}
\{s,\theta\}+(\partial_{\theta}s)^2\text{Re}\{X(s), s\}&=\frac{1}{2}-12\beta_2(\theta),\\
(\partial_{\theta}s)^2\text{Im}\{X(s), s\}&=-4\text{Im}(e^{2i\theta}f(\theta)).
\end{aligned}
\end{equation}
If we know the reparametrization $s(\theta)$, then we can use the boundary data $\{X(s), s\}$ to find $\beta_{2}$ and $f$. The problem remains of how to find the reparametrization $s(\theta)$ given a specific boundary contour. In the following subsection we describe how to find such reparameterization for contours close to circular.

\subsection{Perturbation around the circular contour}
In some very interesting work \cite{Dekel:2015bla}, Dekel applied the above method to contours which are small perturbations of the circular contour. We review those results here since we extend them later using the Shanks transformation and use them to check the results of the new method.
The shape of the contours is taken to be of the form
\begin{equation}
X(\theta)=e^{is(\theta)+\sum_{n=1}^\infty\epsilon^n \xi_n(s(\theta))}
\end{equation}
where $\epsilon$ is the perturbation parameter. Correspondingly, $f(z)$ and $\alpha(z,\bar{z})$ have the expansion:
\begin{equation}
\begin{aligned}
f(z)&=\sum _{n=1}^{\infty}f_n(z)\epsilon^n,\\
\alpha(z,\bar{z})&=\text{ln}(\frac{1}{1-z\bar{z}})+\sum _{n=2}^{\infty}\alpha_n(z,\bar{z})\epsilon^n,
\end{aligned}
\end{equation}
and the correct reparametrization $s(\theta)$ has the expansion
\begin{equation}
s(\theta)=\theta+\sum _{n=1}^{\infty} s_n(\theta)\epsilon^n.
\end{equation}
When $\epsilon=0$, $X(\theta)$, $s(\theta)$, $f(z)$ and $\alpha(z,\bar{z})$ reduce to the results of the circular contour. Given the boundary contour $X(s(\theta))$, one can first calculate the real and imaginary parts of the Schwarzian derivative expressed in terms of the unknown $s_n(\theta)$. Next one expands the LHS of the equations \eqref{schwarzianRI} with the parameter $\epsilon$ and extract $f(\theta)$ and $\beta_2(\theta)$ order by order. Plugging $f(z)$ into the generalized cosh-Gordon equation to solve for $\alpha(z,\bar{z})$ and expanding the solution near the boundary, one gets $\beta_2(\theta)$ which can then be used to compare with the first equation of \eqref{schwarzianRI} to fix $s_n(\theta)$. In the end, one can plug these $s_n(\theta)$ into $f(z)$ and $\alpha(z,\bar{z})$ to calculate the area using \eqref{regarea}.

In \cite{Dekel:2015bla}, this procedure was applied to various contours and the areas were given as a series expansion in terms of $\epsilon$. Here we cite the area formulas for elliptical and symmetric contours $X(s)=e^{is+\epsilon\sin ps}$ \cite{Dekel:2015bla}:
\begin{equation}
\begin{aligned}
\cA_{\text{ellipse}}=&-2\pi-\frac{3\pi\epsilon^2}{4}+\frac{3\pi\epsilon^3}{4}-\frac{237\pi\epsilon^4}{320}+\frac{117\pi\epsilon^5}{160}-\frac{64881\pi\epsilon^6}{89600}\\
&+\frac{64443\pi\epsilon^7}{89600}-\frac{14373577\pi\epsilon^8}{20070400}+\frac{3584953\pi\epsilon^9}{5017600}-\frac{110314688219\pi\epsilon^{10}}{154542080000}\\
&+\frac{22064732579\pi\epsilon^{11}}{30908416000}-\frac{6630907488364381\pi\epsilon^{12}}{9281797324800000}\\
&+\frac{1106373532973931\pi\epsilon^{13}}{1546966220800000}-\frac{40943000996733445243\pi\epsilon^{14}}{57175871520768000000}\\
&+\frac{1952095942839819321\pi\epsilon^{15}}{2722660548608000000}-\frac{157750690929831538029244697\pi\epsilon^{16}}{219774901986388869120000000}\\
&+\frac{19736906966190071806502297\pi\epsilon^{17}}{27471862748298608640000000}\\
&-\frac{801650044535506237372382994066703\pi\epsilon^{18}}{1115068403809909423032238080000000}+\mathcal{O}(\epsilon^{20}), \label{ellipse}
\end{aligned}
\end{equation}
\begin{equation}
\begin{aligned}
\cA_{\text{symmetric,p=2}}=&-2\pi-\frac{3\pi\epsilon^2}{4}+\frac{93\pi\epsilon^4}{20}-\frac{50143\pi\epsilon^6}{4200}
+\frac{510139\pi\epsilon^8}{14400}\\
&-\frac{65754318359\pi\epsilon^{10}}{582120000}+\frac{1195458440855851\pi\epsilon^{12}}{3178375200000}\\
&-\frac{61047851487256409\pi\epsilon^{14}}{47344547250000}+\frac{45707069078388982419341507\pi\epsilon^{16}}{10124976097716480000000
}\\
&-\frac{52566325973037148254959546391187\pi\epsilon^{18}}{3273637646841985463040000000}+\mathcal{O}(\epsilon^{20}),
\end{aligned}
\end{equation}
\begin{equation}
\begin{aligned}
\cA_{\text{symmetric,p=13}}=&-2\pi-1092\pi\epsilon^2+\frac{1660932\pi\epsilon^4}{25}\\
&-\frac{3887594024353\pi\epsilon^6}{570000}+\frac{679687975645852511\pi\epsilon^8}{821712000}\\
&-\frac{2652706006393624451200787779\pi\epsilon^{10}}{24329522800000000}+\mathcal{O}(\epsilon^{12}).
\end{aligned}
\end{equation}
These results will be used to compare with the area calculations in the following sections.

\section{Finding the reparametrization $s(\theta)$}

\begin{figure}[t]
\centering
\includegraphics[trim=0cm 0cm 0cm 3cm, clip=true, width=1.1 \textwidth]{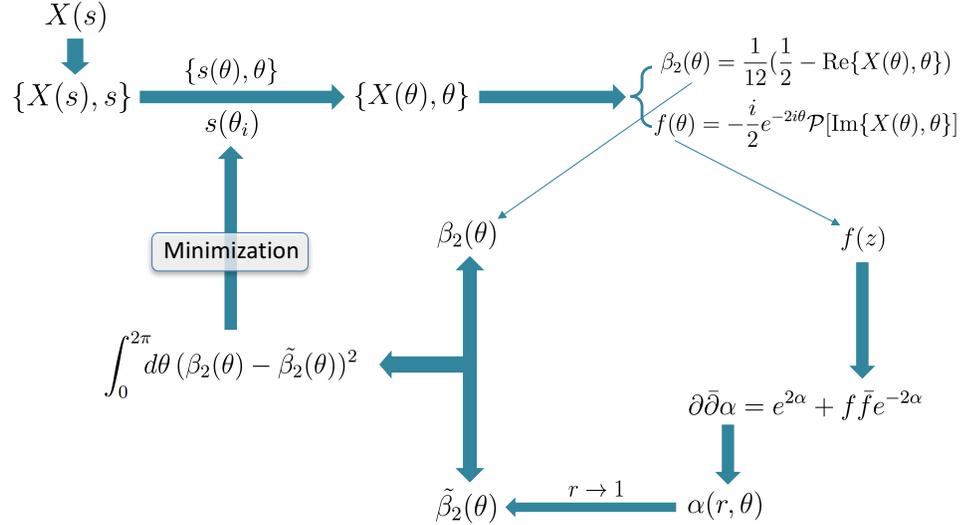}
\caption{The procedure for finding the reparametrization $s(\theta)$. In the expression of $f(\theta)$, $\mathcal{P}$ projects onto positive frequencies.}
\label{procedure}
\end{figure}

Instead of finding the parameterization $s(\theta)$ as a series expansion near a circular contour we can implement a numerical procedure that is in principle defined for any contour. The idea is simple, for a given contour $X(s)$, one proposes a reparametrization $s(\theta)$, and then calculates the Schwarzian derivative of $X(\theta)$ with respect to $\theta$. Thus, a potential value for $\beta_2(\theta)$ and $\text{Im}(e^{2i\theta}f(\theta))$ is found from eq.\eqref{schwarzianRI}. This data can be analytically continued to find $f(z)$ inside the unit disk and then solve the generalized cosh-Gordon equation numerically by a procedure describe in the appendix. Next we expand the resulting $\alpha$ near the boundary and extract the $\beta_2$ according to \eqref{alphaexpand}, and call it $\tilde{\beta}_2$. If $\theta$ is the conformal angle, i.e., $s(\theta)$ is the correct reparametrization, we should have $\beta_2(\theta)=\tilde{\beta}_2(\theta)$. If not we compute the error as
\begin{equation}\label{dbeta2}
B_2[s(\theta)]=\int_0^{2\pi}\!\!d\theta (\beta_2(\theta)-\tilde{\beta}_2(\theta))^2
\end{equation}
 Now we can use standard numerical procedures to find the minimum of $B_2$ as a functional of $s(\theta)$. In practice we define $s(\theta)$ by its values at fixed angles $\theta_j= j\, \frac{2\pi}{M} $, $j=0..M-1$  and use Powell's multidimensional minimization method as described in chapter 10 of \cite{numrec}. The larger the number of interpolating points needed, the more complicated the numerical calculations.
 Once we find the minimum of the function \eqref{dbeta2}, the corresponding $s(\theta)$ will be the best value for the reparametrization of the contour and $B_2$ will be a measure of the error.  The procedure is illustrated in Figure \ref{procedure}. 

 As a test, we applied this minimization procedure for various contours including ellipses, symmetric contours, etc. and found $s(\theta)$ in each case. Then we calculated the areas using \eqref{regarea}. For the contours which have near-circular shapes, we checked the results with the perturbative area formula given in \cite{Dekel:2015bla} and found agreement. For each contour, we also calculated the number $n$ of zeros of $f(z)$ using the formula
\begin{equation}
n=\frac{1}{2\pi}\oint \frac{f'(e^{i\theta})}{f(e^{i\theta})}e^{i\theta}d\theta.
\end{equation}
For the cases where $f(z)$ has no zeros inside the unit disk, we confirm the results of the area calculation with the area formula given in \cite{Kruczenski:2014bla}:
\begin{equation}
\cA_{f}=-2\pi-\bigg|\frac{i}{2}\oint \frac{\text{Re}\{X(\theta), \theta\}-\{\chi,\theta\}}{\partial_{\theta}\ln \chi}d\theta\bigg|,
\label{Areg}
\end{equation}
where
\begin{equation}
\chi(z)=\int^z \sqrt{f}dz.
\end{equation}
For the cases where $f(z)$ has zeros inside the unit disk, another formula for calculating the minimal surface area is given in section \ref{areaforzeros}.  In the following subsections, we describe the results for the reparametrization $s(\theta)$ and the area we obtained for various contours as well as the comparison with calculations using other methods.

\subsection{Ellipse}
The elliptical contour is given by
\begin{equation}
X(s)=\cos(s)+iR\sin(s),
\end{equation}
where $R$ is the ratio of the two axis. As input for the procedure we use the Schwarzian derivative given by
\begin{equation}
\{X(s),s\}=\frac{5-5R^2+(1+R^2)\cos(2s)+4iR\cos(s)\sin(s)}{4(R\cos(s)+i\sin(s))^2}.
\end{equation}
Given the symmetry of the contour, $s(\theta)$ should have rotation and reflection symmetries. Therefore, to look for the reparameterization $s(\theta)$, we only need to minimize the function \eqref{dbeta2} for values of $\theta\in[0,\frac{\pi}{2}]$, which greatly reduces the calculation.

\begin{figure}[H]
\centering
\includegraphics[width=0.8\textwidth]{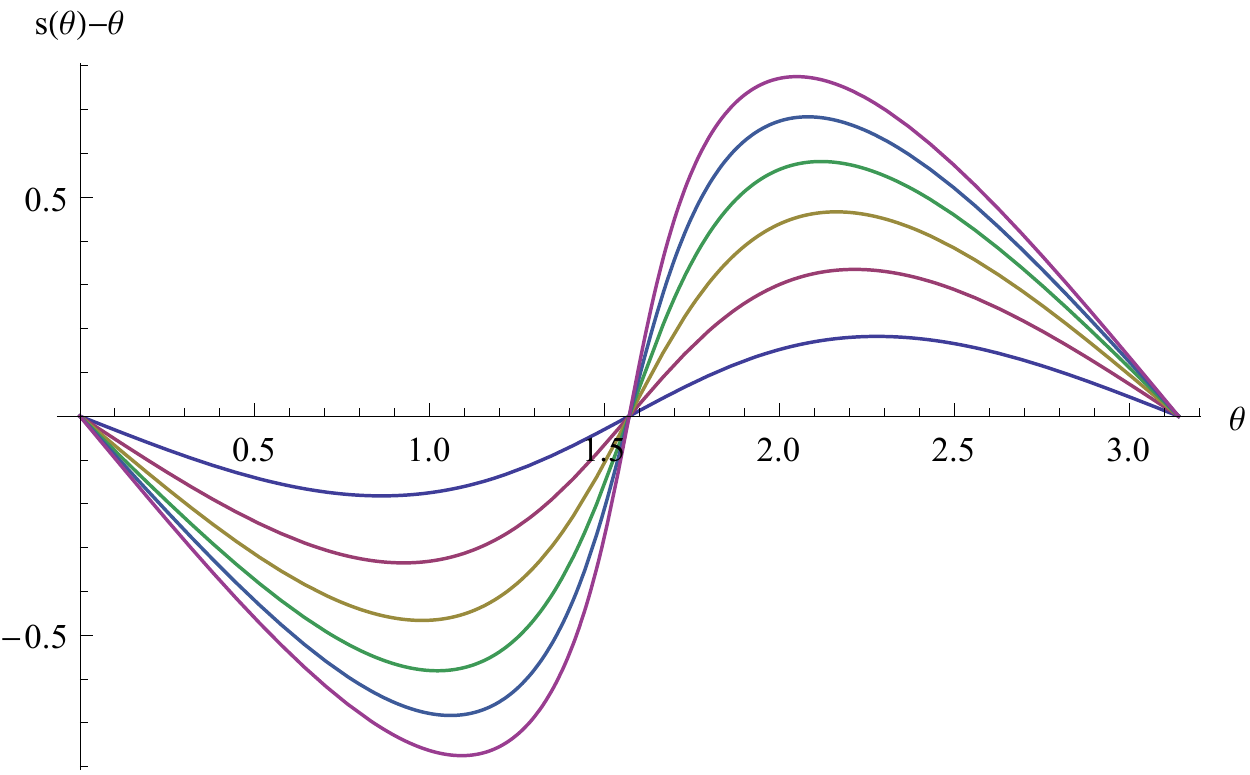}
\caption{The reparametrization function for various ellipses. Here we plot the difference between $s(\theta)$ and $\theta$ for $0<\theta<\pi$. We consider contours with values $1.2\le R\le 2.2$ at $0.2$ intervals.}
\label{ellipserepara}
\end{figure}

\begin{figure}[H]
\centering
\includegraphics[width=0.8\textwidth]{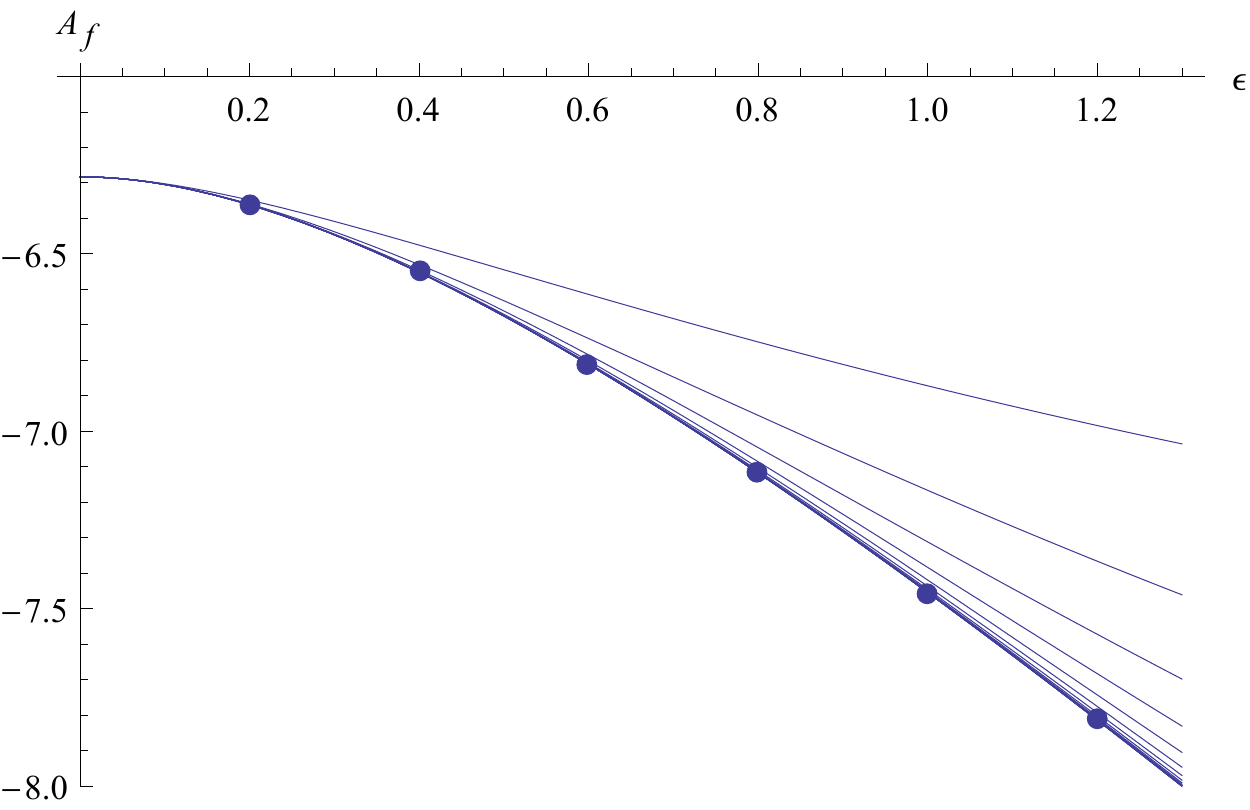}
\caption{The areas of the elliptical contours compared with the perturbative calculations from $\epsilon^2$ up to $\epsilon^{18}$. Notice however that the perturbative results are obtained by using conformal invariance to map $\epsilon$ to $\tilde{\epsilon}=-\frac{\epsilon}{1+\epsilon}$ and computing the better behaved series $\cA_{\text{ellipse}}(\tilde{\epsilon})$. The original series \eqref{ellipse} cannot be used for values of epsilon $\epsilon\gtrsim 0.8$. Such trick is not available for the other contours.}
\label{ellipsearea}
\end{figure}

 We applied the procedure described in this section to elliptical contours with $1.2\le R \le 2.2$ at $0.2$ intervals and find the reparameterizations $s(\theta)$ and the areas. Writing $R=1+\epsilon$, we can compare the areas with the formula given in \cite{Dekel:2015bla}.  See Figure \ref{ellipserepara} and Figure \ref{ellipsearea} where we find  agreement.

\subsection{Symmetric contours}
\begin{figure}[t]
  \begin{subfigure}[b]{0.5\textwidth}
    \includegraphics[width=\textwidth]{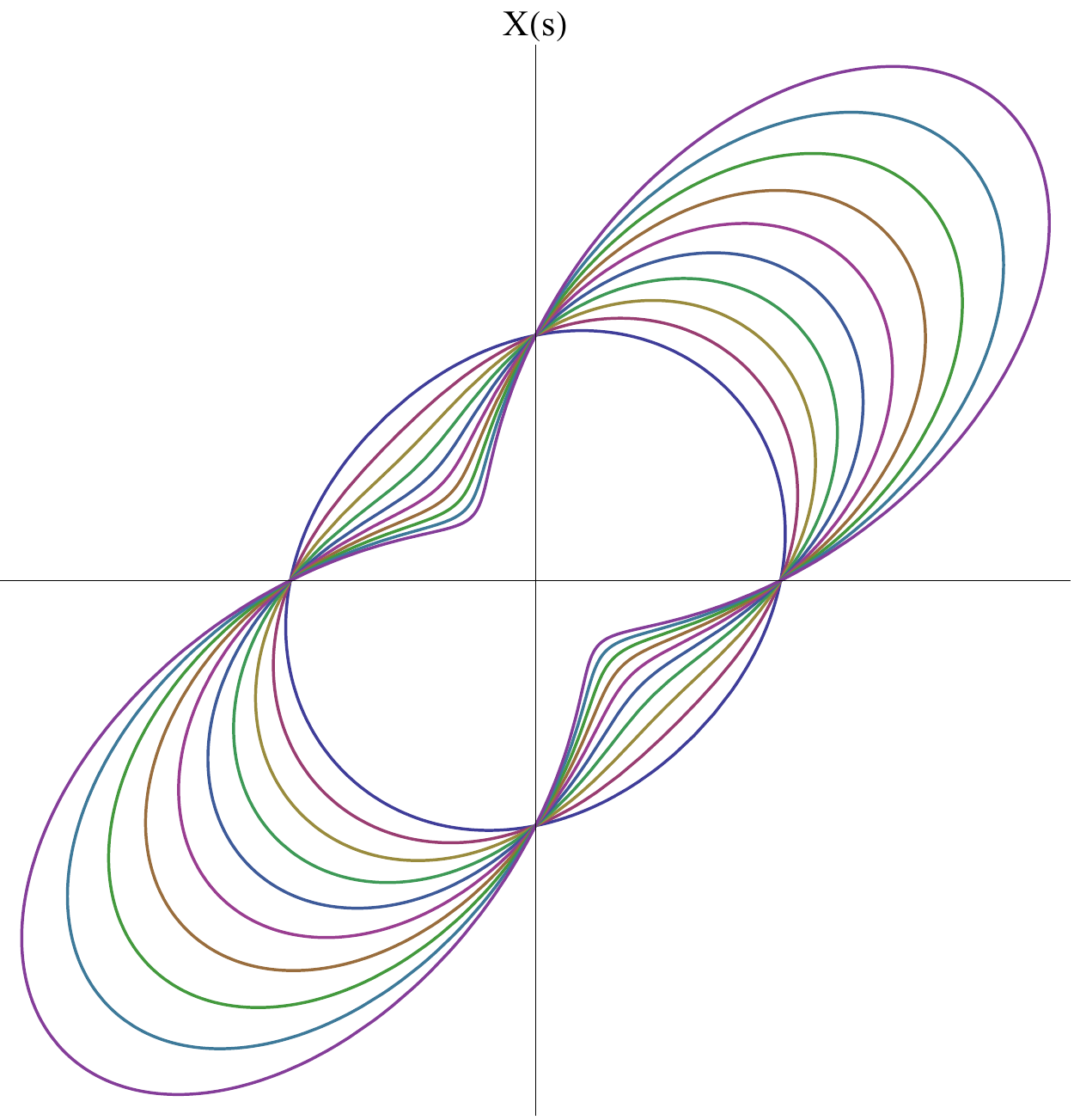}
    \caption{}
    \label{p2}
  \end{subfigure}
  \begin{subfigure}[b]{0.5\textwidth}
    \includegraphics[width=\textwidth]{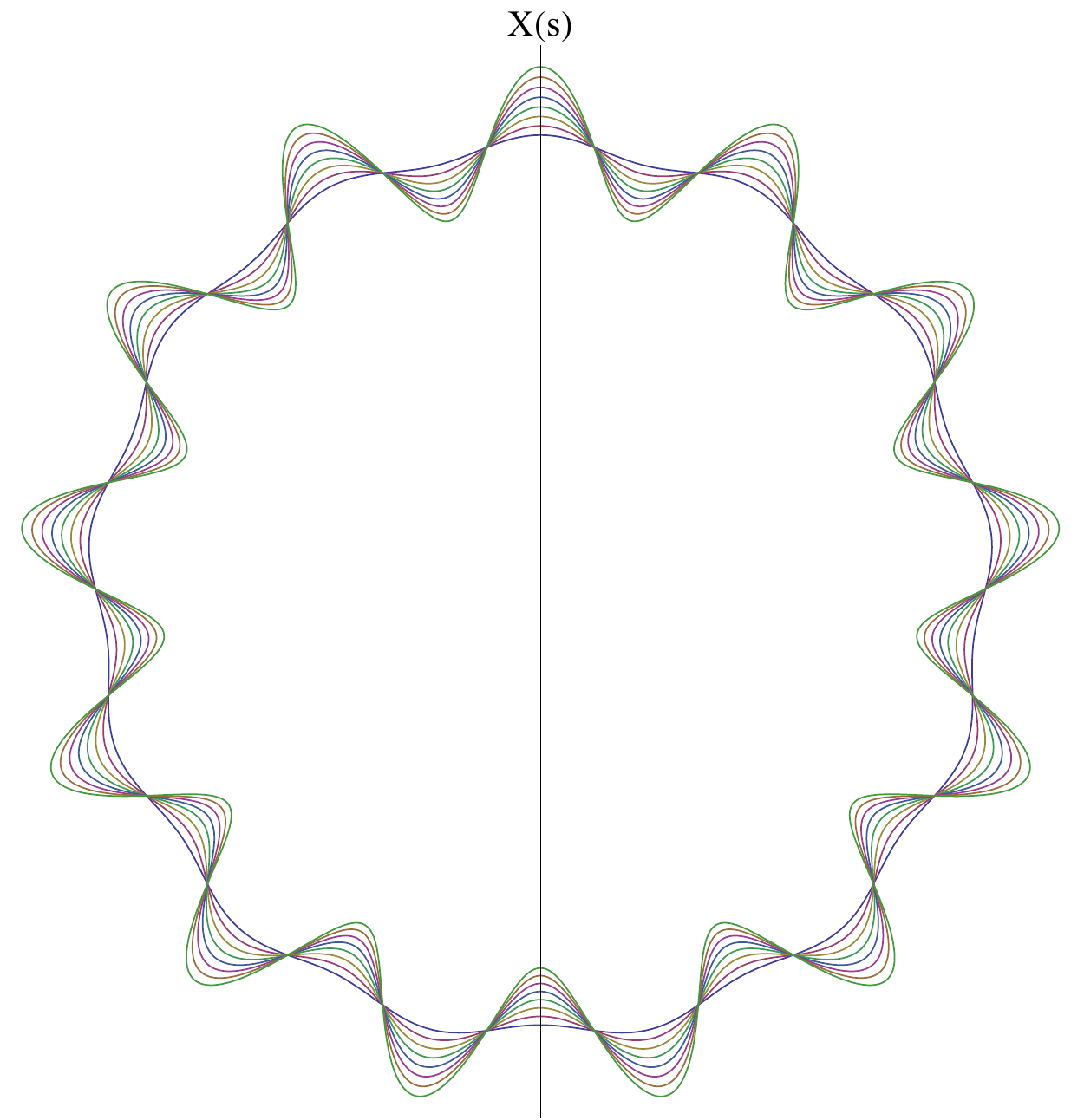}
    \caption{}
    \label{p13}
  \end{subfigure}
  \caption{\ref{p2}: symmetric contours with $p=2$ and $0.1<a<1$ at $0.1$ intervals. \ref{p13}: symmetric contours with $p=13$ and $0.02<a<0.16$ at $0.02$ intervals. They have $\mathbb{Z}_p$ rotational, reflection and inversion symmetries.}
\label{symmetriccontours}
\end{figure}

The symmetric contours in \cite{Dekel:2015bla} are defined by
\begin{equation}
X(s)=e^{i s+a\sin{p s}},
\end{equation}
with $p$ a positive integer (see Figure \ref{symmetriccontours}).
Such contours have $p$-fold rotational symmetry $X\rightarrow e^{\frac{2\pi i}{p}} X$, reflection symmetry $X\rightarrow e^{\frac{i\pi }{p}} \bar{X}$ and inversion symmetry $X\rightarrow X^{-1}$. As a result, the generalized cosh-Gordon equation, and therefore $s(\theta)$ and $\alpha(z,\bar{z})$ have $2p$-fold rotational symmetry. However, $f(z)$ does not have such symmetry. In fact, it can be seen that $f(z)$ has a multiple zero at $z=0$ and if we write $f(z)$ as
\begin{equation}
f(z)=z^{p-2}\tilde{f}(z),
\end{equation}
then $\tilde{f}(z)$ has $2p$ rotational symmetry. When solving for the reparametrization for the symmetric contours, we impose the symmetry condition on the minimization procedure, namely we divide the unit disk into $2p$ wedges and solve the problem on a single wedge.

\begin{figure}[H]
\centering
\includegraphics[width=0.8\textwidth]{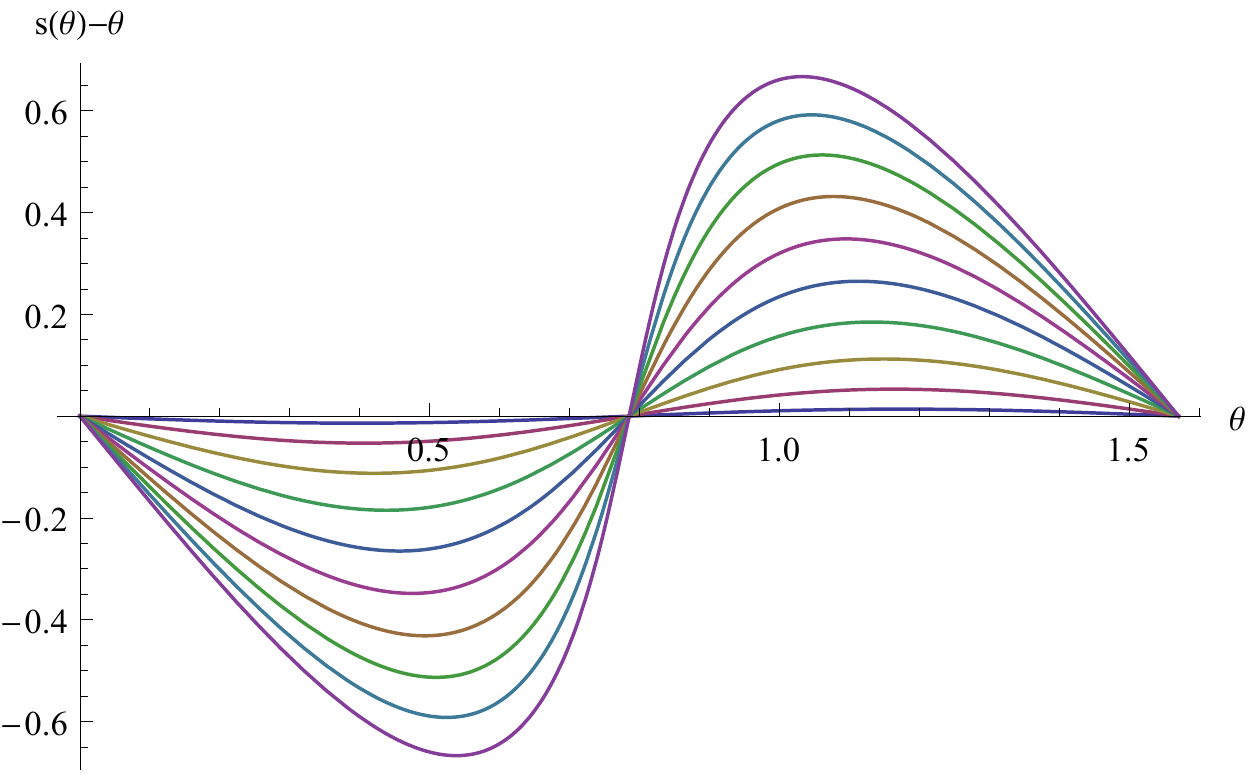}
\caption{The reparametrization functions for symmetric contours with $p=2$. Here we plot the difference between $s(\theta)$ and $\theta$ for $0\le\theta\le\frac{2\pi}{4}$. The contours we consider are $0.1\le a\le 1$ at $0.1$ intervals.}
\label{symmetric2repara}
\end{figure}

\begin{figure}[H]
\centering
\includegraphics[width=0.8\textwidth]{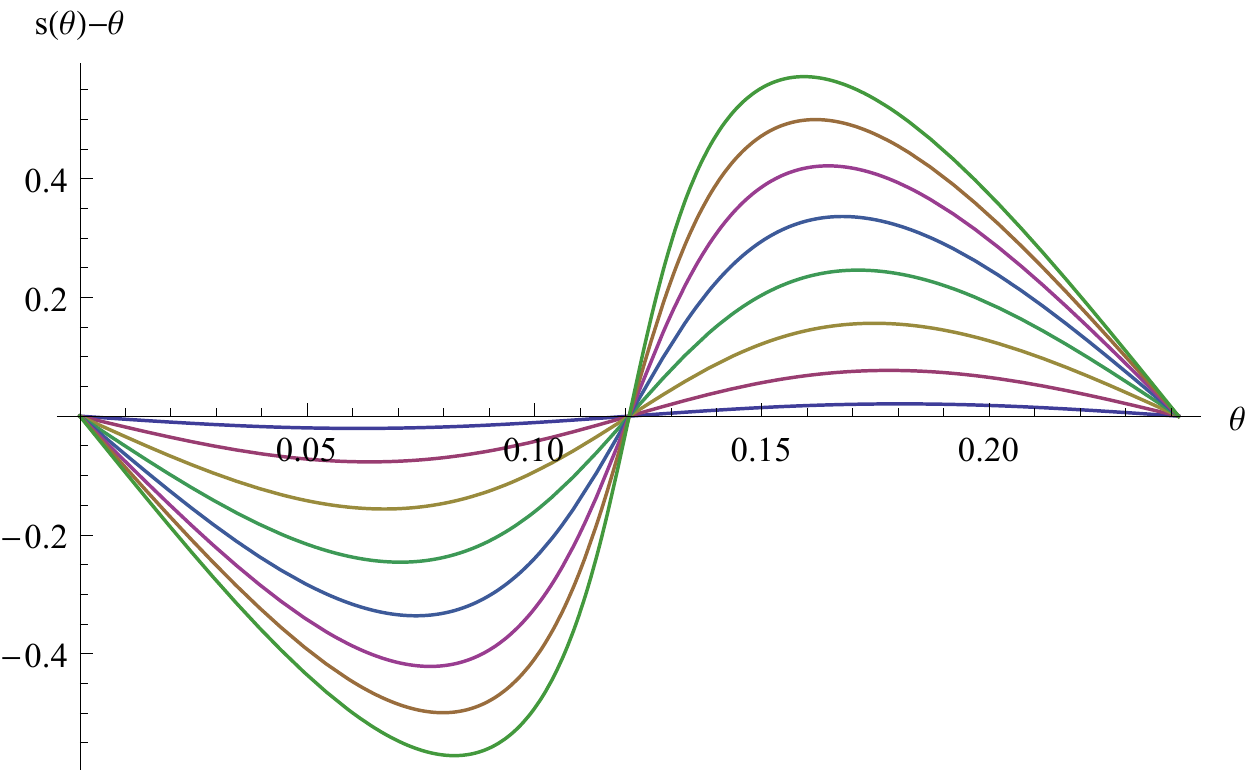}
\caption{The reparametrization functions for symmetric contours with $p=13$. Here we plot the difference between $s(\theta)$ and $\theta$ for the relevant region $0<\theta<\frac{2\pi}{26}$. The contours we consider are $0.02\le a\le  0.16$ at $0.02$ intervals.}
\label{symmetric13repara}
\end{figure}

\begin{figure}[H]
\centering
\includegraphics[width=0.8\textwidth]{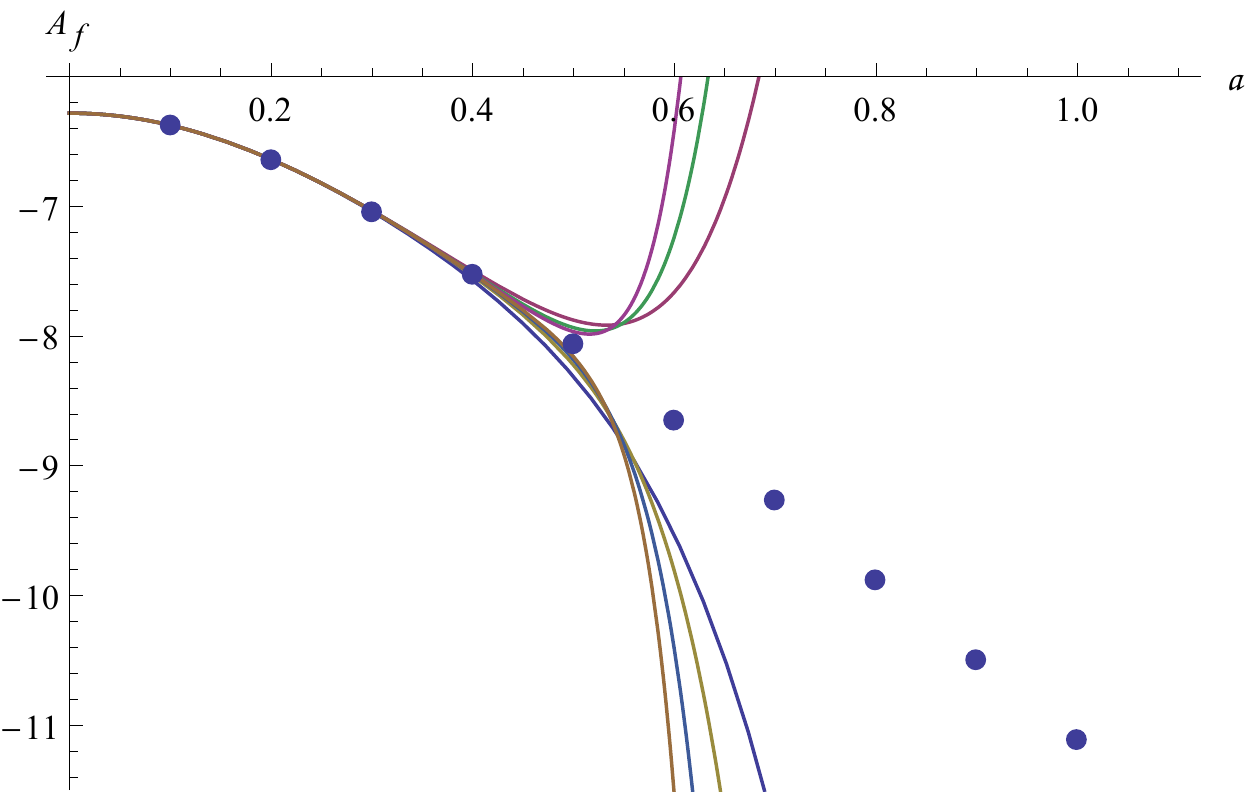}
\caption{The areas of the symmetric contours with $p=2$. We plot a few partial sums (continuous curves) of the perturbative calculations up to the 18th order as well as the results from our calculation (dots), which goes beyond the range of $a$ where the perturbative series converges.}
\label{symmetric2area}
\end{figure}

\begin{figure}[H]
\centering
\includegraphics[width=0.8\textwidth]{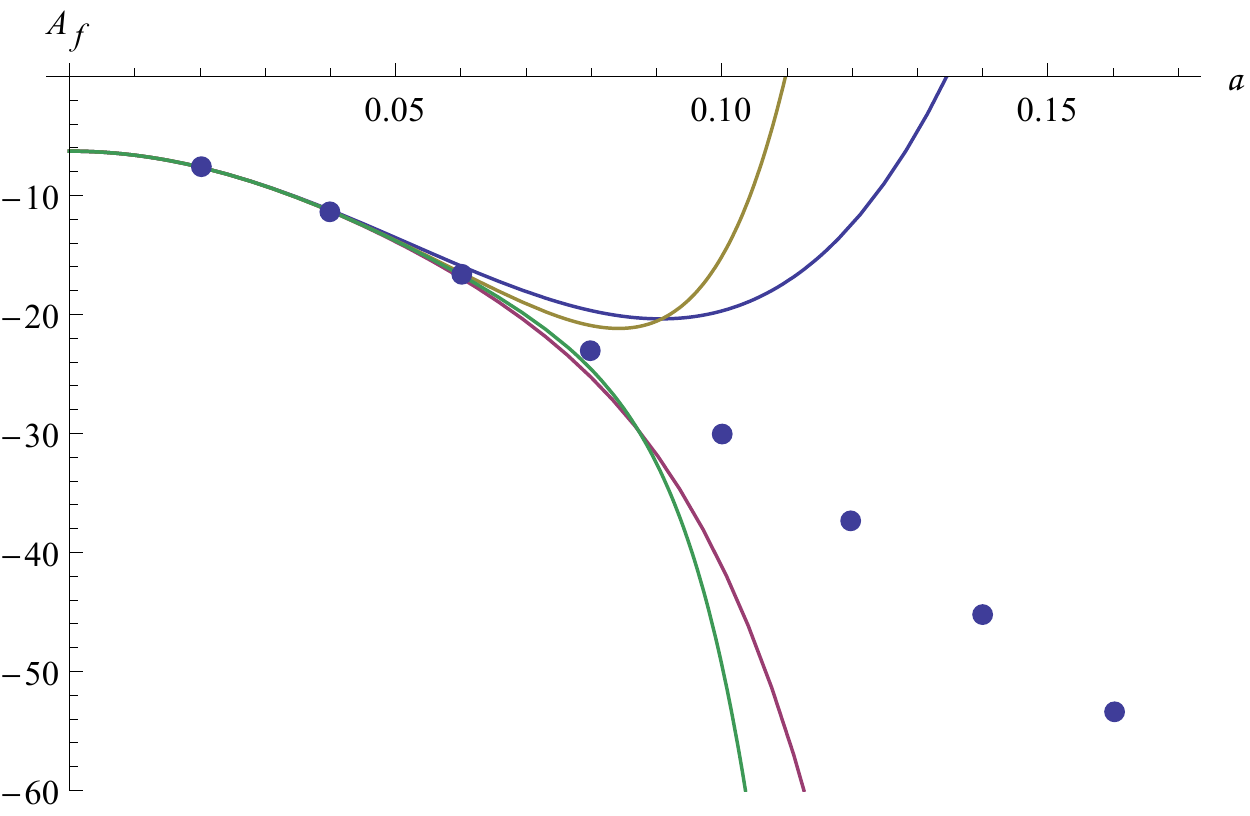}
\caption{The areas of the symmetric contours with $p=13$. We plot a few partial sums (continuous curves) of the perturbative calculations up to the 10th order as well as the results from our calculation (dots), which goes beyond the range of $a$ where the perturbative series converges.}
\label{symmetric13area}
\end{figure}

We perform the calculations for $p=2$ and $p=13$ with different values of $a$. Setting $a=\epsilon$, we compare the areas with the results from \cite{Dekel:2015bla} in the region where the series expansion for the area converges and find agreement.   In Figs. \ref{symmetric2repara} and \ref{symmetric13repara}, we show the reparametrization functions for various symmetric contours and in Fig. \ref{symmetric2area} and \ref{symmetric13area}, we illustrate the comparison between the area calculations.

\section{Shanks transformation}\label{shanks}
As already discussed, in \cite{Dekel:2015bla} the area formula for various shapes is given as a series expansion on the perturbative parameter $\epsilon$. However, the series diverges beyond certain values of $\epsilon$. There are various methods to accelerate the convergence of such a series. We found particularly useful the so called Shanks transformation\cite{Shanks} based on the partial sums of the series
\begin{equation}
A_{N}=\sum_{n=0}^N a_n \epsilon^{n},
\end{equation}
and defined as: 
\begin{equation}
S(A_N)=\frac{A_{N+1}A_{N-1}-A_N^2}{A_{N+1}+A_{N-1}-2A_{N}}.
\end{equation}
By using repeated Shanks transformations for the symmetric contours and the ellipse we observe great improvement of convergence as shown in the following examples.
\subsection{Symmetric contour}
\begin{figure}[H]
\centering
\includegraphics[width=0.8\textwidth]{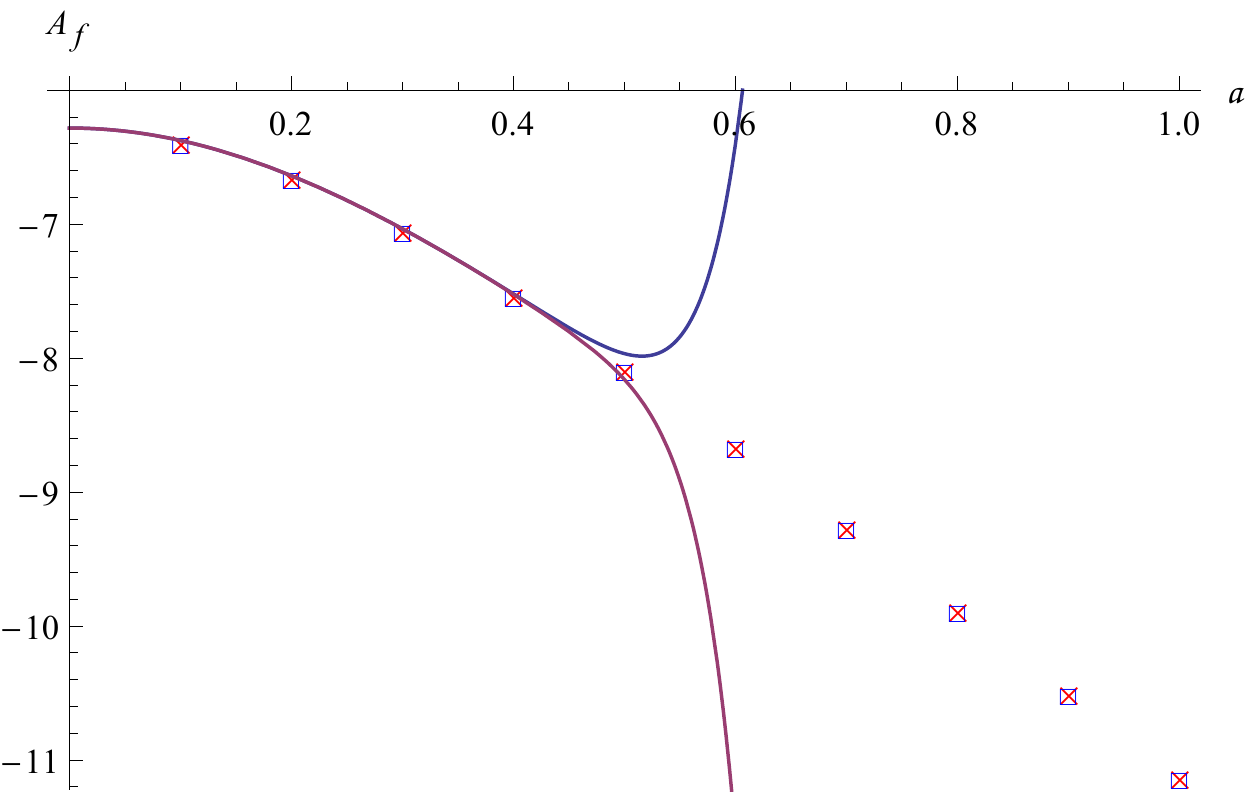}
\caption{Comparison of different area calculations for symmetric contours with $p=2$. {\color{red}{$\times$}} indicates the areas calculated by finding the reparametrization, and {\color{blue}{$\square$}} indicates the areas calculated through Shanks transformation of the perturbative expansion. The results agree.}
\label{areacomparison}
\end{figure}

For symmetric contours,  the series expansion of the area has only even powers of $\epsilon$. We therefore write the area series as
\begin{equation}
\cA_{sym,N}=\sum_n^N a_{sym,n} \epsilon^{2n},
\end{equation}
and perform Shanks transformation on it. As can be seen in Figure \ref{symmetric2area} and \ref{symmetric13area}, the series diverges at around $\epsilon \sim 0.4$ and $\epsilon \sim 0.13$ for countours with $p=2$ and $p=13$ respectively. After the Shanks transformation with the coefficients given in \cite{Dekel:2015bla}, we manage to find the areas for the values of $a$ used in the previous section and find good agreement. See Table \ref{shanks0.7}, and figure \ref{areacomparison} for the comparison.
\begin{center}
\begin{tabular}{l*{5}{c}}
$N$ & $A_N$ & $S(A_N)$ & $S^2(A_N)$ & $S^3(A_N)$ & $S^4(A_N)$\\
\hline
\\
1 & -10.9013 &  &   &  & \\
2 & -7.39385 & -9.34802 &   &   &  \\
3 & -11.8065 & -9.19200 & -9.25671 &   &  \\
4 & -5.39056 & -9.30258 & -9.24927 & -9.25187 &  \\
5 & -15.4146 & -9.19965 & -9.25326 & -9.25161 &  -9.25169\\
6 & 0.940615 & -9.31152 & -9.25044 & -9.25173 &  \\
7 & -26.5334 & -9.17697 & -9.25281 &   &  \\
8 & 20.59769 & -9.35077 &   &   &  \\
9 & -61.5493 &  &  &   &  \\
\end{tabular}
\captionof{table}{Shanks transformation of the area series for $p=2$, $a=0.7$. The result obtained by finding the reparametrization which is shown in Figure \ref{symmetric2area} is $\cA_{f}(a=0.7)=-9.25174$.}
\label{shanks0.7}
\end{center}

\subsection{Ellipse}
\begin{figure}[H]
\centering
\includegraphics[width=0.8\textwidth]{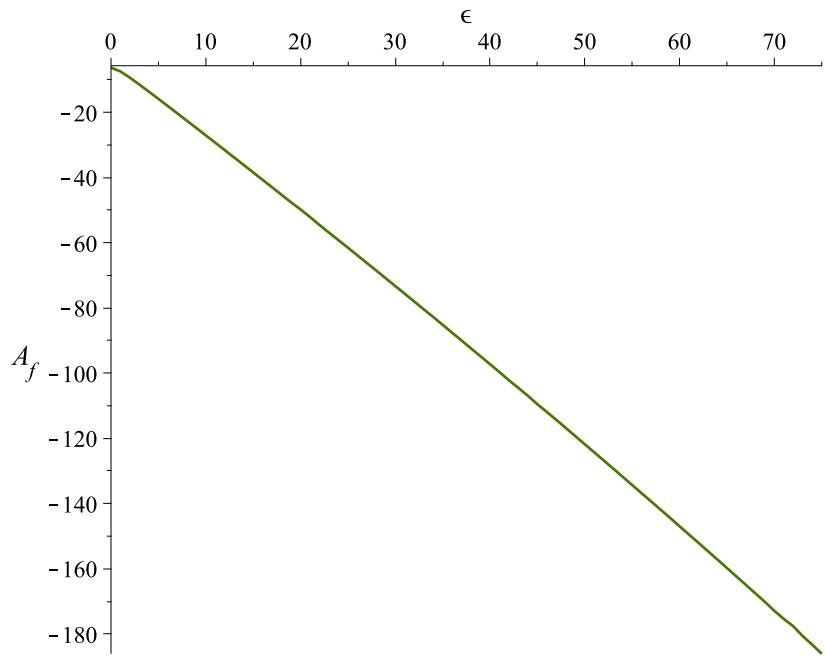}
\caption{Plots of minimal surface area for an ellipse boundary as a function of $\epsilon=R-1$  after performing nine Shanks transformations of the series expansion \eqref{ellipse} in terms of $\tilde{\epsilon}=-\frac{\epsilon}{1+\epsilon}$. See Table \ref{shank10}.}
\label{shanksellipse}
\end{figure}

For elliptical contours, Dekel used conformal symmetry to relate an ellipse with perturbative parameter $\epsilon$ to one with $\tilde{\epsilon}=-\frac{\epsilon}{1+\epsilon}$. While $0<\epsilon<\infty$, we have $-1<\tilde{\epsilon}<0$. Therefore, we can consider the areas for ellipses with $-1<\tilde{\epsilon}<0$ where the perturbative formula has better convergence. However, the approximation fails for $\epsilon\gtrsim 4$. We apply Shanks transformation on the area formula for the ellipse with $-1<\tilde{\epsilon}<0$. The convergence of the series accelerates drastically. In Table \ref{shank10}, we show the acceleration of convergence for one elliptical contour beyond the range of convergence of the original formula ($\epsilon=10$). In fact we see convergence up to $\epsilon \sim 100$. We plot those results in Fig.\ref{shanksellipse}.

\begin{sidewaystable}
	\begin{tabular}{*{11}{r}}
		$N$ & $A_N$ & $S(A_N)$ & $S^2(A_N)$ & $S^3(A_N)$ & $S^4(A_N)$ & $S^5(A_N)$ & $S^6(A_N)$ & $S^7(A_N)$& $S^8(A_N)$& $S^9(A_N)$\\
		\hline		\\
	 	 0 &   -6.2832 &               &                &               &               &               &               &               &              &               \\
	 	 1 &   -6.2832 &   -6.2832 &                &               &               &               &               &               &              &               \\
		 2 &   -8.2305 & -27.7031 &  -25.7380 &               &               &               &               &               &              &               \\
		 3 & -10.0007 & -25.5395 &  -25.5175 & -25.5365 &               &               &               &               &              &               \\
		 4 & -11.5899 & -25.5177 &  -25.5383 & -25.5174 & -25.5365 &               &               &               &              &               \\
		 5 & -13.0163 & -25.8864 &  -45.5321 & -36.0704 & -31.2791 & -28.8814 &               &               &              &               \\
		 6 & -14.3004 & -26.2483 &  -27.5703 & -27.3039 & -27.1627 & -26.8718 & -26.9845 &               &              &               \\
		 7 & -15.4599 & -26.5324 &  -27.3078 & -27.1649 & -26.8919 & -26.9915 & -27.1289 & -27.0797 &              &               \\
		 8 & -16.5095 & -26.7404 &  -27.2153 & -27.0728 & -27.0495 & -27.0552 & -27.0542 & -27.0547 &-27.0548 &               \\
		 9 & -17.4614 & -26.8850 &  -27.1592 & -27.0542 & -27.0550 & -27.0542 & -27.0547 & -27.0548 &-27.0550 & -27.0553 \\
		10 & -18.3260 & -26.9797 & -27.1226 & -27.0550 & -27.0541 & -27.0552 & -27.0546 & -27.0546 &-27.0550 &               \\
		11 & -19.1121 & -27.0366 & -27.0989 & -27.0600 & -27.0494 & -27.0538 & -27.0552 & -27.0546 &              &               \\
		12 & -19.8272 & -27.0664 & -27.0842 & -27.0693 & -27.1378 & -27.1017 & -27.0858 &               &              &               \\
		13 & -20.4780 & -27.0775 & -27.0768 & -27.0775 & -27.0768 & -27.0779 &               &               &              &               \\
		14 & -21.0704 & -27.0767 & -27.0776 & -27.0767 & -27.0779 &               &               &               &              &               \\
		15 & -21.6097 & -27.0690 & -27.0943 & -27.0742 &               &               &               &               &              &               \\
		16 & -22.1004 & -27.0578 & -27.1938 &               &               &               &               &               &              &               \\
		17 & -22.5470 & -27.0457 &               &               &               &               &               &               &              &               \\
		18 & -22.9532 &               &               &               &               &               &               &               &              &               \\
	\end{tabular}
	\captionof{table}{Shanks transformation of the area series for ellipse with $\epsilon=R-1=10$. We used $\tilde{\epsilon}=-\frac{\epsilon}{1+\epsilon}$ and Shanks transform the series for $\tilde{\epsilon}=-\frac{10}{11}$. The area shows good convergence after nine Shanks transformations.}
	\label{shank10}
\end{sidewaystable}

\section{Area formula for $f(z)$ with zeros}\label{areaforzeros}

\begin{figure}[H]
\centering
\includegraphics[width=0.8\textwidth]{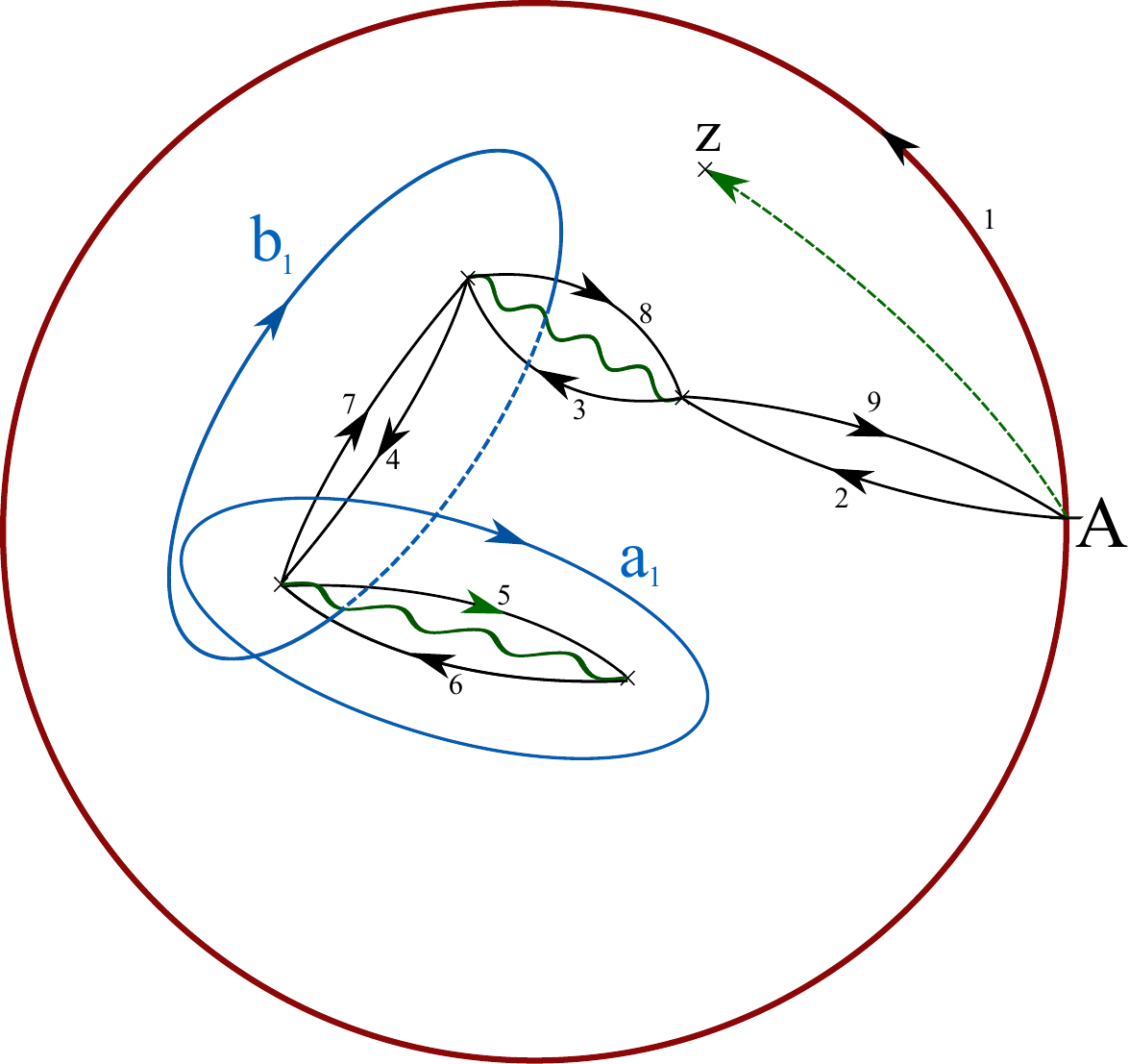}
\caption{If the holomorphic function $f(z)$ has zeros, the formula for the area gets an extra contribution from integrals around the non-trivial contours.}
\label{fzeros}
\end{figure}

If we define the one forms $j$, $w$, and function $\chi$ on the world-sheet \cite{Alday:2010vh,Irrgang:2015txa}
 \beqa
 j &=& 4 f \sqrt{\bar{f}}  e^{-2\alpha} dz + \frac{2}{\sqrt{f}} \left(\bar{\partial}^2\alpha-(\bar{\partial}\alpha)^2\right) d\bar{z} \\
 \chi &=& \int_{A}^{z} w, \ \ \ w=\sqrt{\bar{f}} d\bar{z}
 \eeqa
where $A$ is any point on the disk, usually at the boundary, then the current $j$ satisfies $dj=0$ as follows from the generalized cosh-Gordon equation \eqref{coshGordon}.
With these definitions, the area can be written as
 \beq
  \cA_f+ 2\pi = -\int dz\,d\bar{z}\ e^{-2\alpha} f\bar{f} = \int j\wedge w = \int j\wedge d\chi = -\int d(\chi j) = -\oint \chi j
 \eeq
 Since $\chi$ is uniquely defined only in a simple connected domain, when $\sqrt{f(z)}$ has cuts, namely when $f(z)$ has zeros, such domain has to go around the cuts. 
For example in fig.\ref{fzeros} we show how to cut the disk following the black lines along a contour labeled by successive segments $1$ to $9$. The lines are separated for clarity but lines $2,9$ actually overlap, as well as $4,7$ etc.    
A straight-forward calculation leads to 
\beqa
 \cA_f+ 2\pi &=& -\frac{i}{4}\left[ \sum_\ell \left(\oint_{a_\ell}w\oint_{b_\ell} j-\oint_{b_\ell} \omega \oint_{a_\ell} j \right)  - \oint_{1} \omega\oint_1 j \right. \\
  &&\ \ \left. + 2\left(\oint_1 \omega\int_2j-\int_2 \omega\oint_1 j\right)+2\oint_1(\int_{A}^{z}\omega)j\right]
\eeqa
where $a_\ell,b_{\ell}$ is a basis of cycles for the disk with cuts and $1$ is the boundary of the disk. The path $2$ connects one cut to the boundary as in the Fig.\ref{fzeros}. This formula is similar to the one known for light-like Wilson loops \cite{Alday:2010vh} except that it contains a contribution from the boundary of the disk. Recall that in the case of \cite{Alday:2010vh} the world-sheet was the whole plane and there was no contribution from infinity. We checked that when $f(z)$ has no zeros it reduces to formula \eqref{Areg} and also we checked numerically the validity of the formula for some examples. It should be noted that this formula requires knowing the values of $\alpha$ and $f$ inside the disk in which case it might be more convenient to directly use the definition \eqref{regarea} as we did previously.

\section{Conclusions}

 In this work we have shown that the integrability ideas to find minimal area surfaces discussed in \cite{Kruczenski:2014bla,Dekel:2015bla} can be implemented numerically. The method is in principle valid for any contour but in practice it becomes numerically difficult if the contour is not reasonably smooth.
 Although one can also try to find the minimal surface by direct minimization of the area functional, using integrability has some advantages. First the method is manifestly invariant under global conformal transformations, second it reconstructs the Pohlmeyer analytic function $f(z)$ 
and therefore it makes it easier to obtain the $\lambda$-deformed contours. More generically, the idea is to understand the role of integrability in the minimal area problem rather than find actual solutions. Along these lines the idea of $\lambda$-deformations \cite{Ishizeki:2011bf,Kruczenski:2014bla,Dekel:2015bla} or "master" symmetry \cite{Klose:2016uur,Klose:2016qfv} seems quite powerful.

\section{Acknowledgments}

 We are grateful to A. Dekel, T. Klose, S. Komatsu, J. Toledo and P. Vieira for various discussions and suggestions on the minimal area problem.  This work was supported in part by DOE through grant DE-SC0007884. 

\appendix
\renewcommand{\theequation}{\Alph{section}.\arabic{equation}}
\section{Solving the generalized cosh-Gordon equation using spectral methods}\label{solvingcoshgordon}
In this section, we describe the numerical method for solving the generalized cosh-Gordon equation \eqref{coshGordon} for arbitrary $f(z)$. The method was used in the procedure of finding the reparametrization $s(\theta)$ described in the main text.

From \eqref{alphaexpand}, we can see that $\alpha(z,\bar{z})$ diverges near the boundary. Therefore, we define
\begin{equation}
\tilde{\alpha}(z,\bar{z})=\alpha(z,\bar{z})+\ln(1-r^2).
\end{equation}
It has the expansion ($\xi=1-r^2$)
\begin{equation}\label{alphatildeexpansion}
\tilde{\alpha}(z,\bar{z})\simeq \beta_2(\theta)(1+\xi)\xi^2+O(\xi^4),
\end{equation}
near the boundary and is finite. Thus, we solve the differential equation for $\tilde{\alpha}$ instead.

 Using the spectral method described in \cite{spectral} requires putting a grid on the unit disk on the $z$ plane and defining functions by their values on the nodes of the grid. Once this is done, the differentiation along the radial and angular directions are calculated through multiplication by differentiation matrices. Although this is standard we had to implement some simple modifications to account for the fact that some functions have double zeros at the boundary of the disk $r=\pm1$. For that reason we briefly discuss the actual implementation used in the paper. 

In the angular direction $\theta$, we choose a periodic grid of $N_{\theta}$ points. The spacing of the grid is
\begin{equation}
h=\frac{2\pi}{N_{\theta}}.
\end{equation}
For a periodic function $p(\theta)$, given its values on the grid $p_i=p(\theta_i)$ with $i=1,2,...N_\theta$, the function can be interpolated as
\begin{equation}
p(\theta)=\sum_{i=1}^{N_\theta}p_i S_{N_\theta}(\theta-\theta_i).
\end{equation}
Here $S_{N_\theta}$ is the periodic sinc function
\begin{equation}
S_{N_\theta}(x)=\frac{\sin(\pi x/h)}{(2\pi/h)\tan(x/2)}.
\end{equation}
The differentiation matrices on the angular grid are obtained by taking derivatives of $S_{N_\theta}$ and evaluating them at the grid points. The first order differentiation matrix is
\begin{equation}
D_{N_\theta}=\begin{pmatrix}
0& & & &-\frac{1}{2}\cot\frac{1h}{2}\\
-\frac{1}{2}\cot\frac{1h}{2}&\ddots& &\ddots& \frac{1}{2}\cot\frac{2h}{2}\\
\frac{1}{2}\cot\frac{2h}{2}& &\ddots& &-\frac{1}{2}\cot\frac{3h}{2}\\
-\frac{1}{2}\cot\frac{3h}{2}&\ddots& &\ddots&\vdots\\
\vdots& &\ddots& &\frac{1}{2}\cot\frac{1h}{2}\\
\frac{1}{2}\cot\frac{1h}{2}& & & & 0
\end{pmatrix},
\end{equation}
i.e., the $(i,j)$ element is given by
\begin{equation}
S'_{N_\theta}(\theta_i-\theta_j)=
\begin{cases}
   0,& i-j=0\quad(\text{mod} N_\theta),\\
   \\
   \frac{1}{2}(-)^{(i-j)}\cot\frac{(i-j)h}{2},& i-j\neq0\quad(\text{mod} N_\theta).
\end{cases}
\end{equation}
For the second derivative, the differentiation matrix is
\begin{equation}
D_{N_\theta}^{(2)}=\begin{pmatrix}
& \ddots& &\vdots& & & \\
& \ddots& &-\frac{1}{2}\csc^2(\frac{2 h}{2})& & & \\
& & &\frac{1}{2}\csc^2(\frac{1 h}{2})& & & \\
& \cdots& &-\frac{\pi^2}{3 h^2}-\frac{1}{6}& &\cdots& \\
& & &\frac{1}{2}\csc^2(\frac{1 h}{2})& & & \\
& & &-\frac{1}{2}\csc^2(\frac{2 h}{2})& &\ddots& \\
& & &\vdots& &\ddots& \\
\end{pmatrix},
\end{equation}
with the $(i,j)$ elements given by
\begin{equation}
S''_{N_\theta}=
\begin{cases}
   -\frac{\pi^2}{3 h^2}-\frac{1}{6},& i-j=0\quad(\text{mod} N_\theta),\\
   \\
   -\frac{(-)^{(i-j)}}{2\sin^2((i-j) h/2)},& i-j\neq0\quad(\text{mod} N_\theta).
\end{cases}
\end{equation}

In the radial direction, we take Chebyshev points
\begin{equation}
r_i=\cos(i\pi/N_r),\quad i=0,1,...,N_r.
\end{equation}
Notice that $r_i\in(-1,1)$. For a function $q(r)$ in the radial direction, if we know the values at Chebyshev points $q_i=q(r_i)$, we can interpolate the function as
\begin{equation}
q(r)=\sum_{i=0}^{N_r}q_iP_i(r),
\end{equation}
where $P_i(r)$ is given by
\begin{equation}
P_i(r)=\frac{\displaystyle\prod_{j\ne i}^{N_r} (r-r_j)}{\displaystyle\prod_{j\ne i}^{N_r} (r_i-r_j)}.
\end{equation}
The differentiation matrix $D_{N_r}$ is given by the derivatives of the interpolation function, $(D_{N_r})_{ij}=P_i'(r_j)$ and $(D_{N_r}^{(2)})_{ij}=P_i''(r_j)$. Specifically, the entries are
\begin{equation}
\begin{aligned}
&(D_{N_r})_{00}=\frac{2N_r^2+1}{6},\qquad(D_{N_r})_{N_rN_r}=-\frac{2N_r^2+1}{6},\\
&(D_{N_r})_{jj}=\frac{-r_j}{2(1-r_j^2)},\qquad j=1,...,N_r-1,\\
&(D_{N_r})_{ij}=\frac{c_i}{c_j}\frac{(-)^{i+j}}{(r_i-r_j)},\qquad i\ne j,\qquad i,j=0,...,N_r,\\
\end{aligned}
\end{equation}
where
\begin{equation}
c_i=\begin{cases}
2,\quad i=0, N_r,\\
1,\quad \mbox{otherwise}.
\end{cases}
\end{equation}

For solving \eqref{coshGordon}, since $\tilde{\alpha}(z,\bar{z})$ has double zeros near the boundary $r\to 1$ as evident from \eqref{alphatildeexpansion}, we will consider the interpolation and differentiation of radial functions with double zeros on the boundary. The interpolation function then takes the form
\begin{equation}
\hat{P}_i(r)=\frac{(1-r^2)^2}{(1-r_i^2)^2}P_i(r).
\end{equation}
Correspondingly, we need to modify the differentiation matrices. The first and second order differentiation matrices will therefore be given by $(\hat{D}_{N_r})_{ij}=\hat{P}_i'(r_j)$ and $(\hat{D}_{N_r}^{(2)})_{ij}=\hat{P}_i''(r_j)$. It is easy to derive the following relations:
\begin{equation}
\begin{aligned}
&(\hat{D}_{N_r})_{ij}=\frac{(1-r_j^2)}{(1-r_i^2)}(D_{N_r})_{ij},\quad (\hat{D}_{N_r})_{jj}=-\frac{2r_j}{(1-r_j^2)}+(D_{N_r})_{jj},\\
&(\hat{D}_{N_r}^{(2)})_{ij}=-\frac{4r_j}{(1-r_i^2)}(D_{N_r})_{ij}+\frac{(1-r_j^2)}{(1-r_i^2)}(D_{N_r}^{(2)})_{ij},\\
&(\hat{D}_{N_r}^{(2)})_{jj}=-\frac{2}{1-r_j^2}P_j(r_j)-\frac{4r_j}{(1-r_j^2)}(D_{N_r})_{jj}+(D_{N_r}^{(2)})_{jj}.
\end{aligned}
\end{equation}

With the differentiation matrices in hand, we can define the following linear operator
\beq
\hat{L} [\tilde{\alpha}] = \partial_r^2\tilde{\alpha}+\frac{1}{r}\partial_r\tilde{\alpha}+\frac{1}{r^2}\tilde{\alpha}-\frac{8\tilde{\alpha}}{(1-r^2)^2}
\eeq
and write the generalized cosh-Gordon equation as
\beq\label{pde1}
\hat{L}[\tilde{\alpha}] =R(\tilde{\alpha})
\eeq
where $R(\alpha)$ is the non-linear function 
\beq
R(\tilde{\alpha}) = \frac{4}{(1-r^2)^2}(e^{2\tilde{\alpha}}-1)+4f\bar{f}(1-r^2)^2e^{-2\tilde{\alpha}}-\frac{8\tilde{\alpha}}{(1-r^2)^2}
\eeq
Notice that the term $-\frac{8\tilde{\alpha}}{(1-r^2)^2}$ appears on both sides of the equation and can be canceled but keeping it results in a well behaved iteration procedure:
\beq
 \tilde{\alpha}^{[n+1]} = \hat{L}^{-1}[R(\tilde{\alpha}^{[n]})].
 \label{iteration}
\eeq
 Following \cite{spectral}, we implement the iterative procedure by using a grid on
\begin{equation}
r\in(-1,1),\qquad \theta\in(0,2\pi)
\end{equation}
and define
\begin{equation}
N_{r2}=\frac{N_r-1}{2},\quad N_{\theta2}=\frac{N_{\theta}}{2}
\end{equation}
with $N_r$ odd. Therefore we will be solving in a grid with $(N_r-1)N_\theta\times(N_r-1)N_\theta$ dimensional differentiation matrices. However, in the end, we only keep the $N_{r2}\times N_\theta$ dimensional solution, which correspond to $r\in(0,1)$ and $\theta\in(0,2\pi)$. This is done by writing the Laplacian operator as
\begin{equation}
L=(D_1+RE_1)\otimes\begin{pmatrix}
I & 0\\
0 & I
\end{pmatrix}+(D_2+RE_2)\otimes\begin{pmatrix}
0 & I\\
I & 0
\end{pmatrix}+R^2\otimes D_{N_\theta}^{(2)},
\end{equation}
where $D_1$, $E_1$ are the upper left block of $D_{N_r}^{(2)}$, $D_{N_r}$ respectively and $D_2$, $E_2$ are the upper right block of $D_{N_r}^{(2)}$, $D_{N_r}$ respectively. $I$ is $N_{\theta2}\times N_{\theta2}$ dimensional identity matrix and $R$ is given by
\begin{equation}
R=\text{diag}(r_i^{-1}),\quad i=1,2,..., N_{r2}.
\end{equation}
For more details, see chapter 11 of \cite{spectral}. Using the spectral method almost any simple initial guess for $\tilde{\alpha}$ converges fast to the solution through the iteration \eqref{iteration}.



\end{document}